\documentclass[modern]{aastex61}
\usepackage{url}

\def\beq{\begin{equation}}
\def\eeq{\end{equation}}
\def\bey{\begin{eqnarray}}
\def\eey{\end{eqnarray}}

\def\Mpc{\,{\rm Mpc}}
\def\mpc{\, h^{-1}{\rm {Mpc}}}
\def\kpc{\, h^{-1}{\rm {kpc}}}
\def\kms{\,{\rm {km\, s^{-1}}}}

\def\msun{\, h^{-1}{\rm M_\odot}}

\def\Mvir{M_{\rm vir}}

\def\mp{m_{\rm p}}
\def\Vmax{V_{\rm max}}
\def\V200{V_{\rm 200}}
\def\r200{r_{\rm 200}}

\def\Mpeak{M_{\rm peak}}
\def\M0{M_{\rm 0}}

\def\ztp{z_{\rm tp}}
\def\zpeak{z_{\rm peak}}
\def\zf{z_{\rm f}}
\def\zn{a_{\rm nf}}

\shorttitle{Bimodal Formation Time Distribution}
\shortauthors{Shi et al.}

\begin{document}

\title{Bimodal Formation Time Distribution for Infall Dark Matter Halos}

\author{Jingjing Shi}
\affiliation{Key Laboratory for Research in Galaxies and Cosmology, Department of Astronomy, University of Science and Technology of China, Hefei, Anhui 230026, China}
\affiliation{SISSA, Via Bonomea 265, I-34136 Trieste, Italy}
\affiliation{Kavli Institute for Astronomy and Astrophysics, Peking University, Beijing 100871, China}

\author{Huiyuan Wang}
\affiliation{Key Laboratory for Research in Galaxies and Cosmology, Department of Astronomy, University of Science and Technology of China, Hefei, Anhui 230026, China}
\affiliation{School of Astronomy and Space Science, University of Science and Technology of China, Hefei 230026, China}

\author{H.J. Mo}
\affiliation{Department of Astronomy, University of Massachusetts, Amherst MA 01003-9305, USA}
\affiliation{Astronomy Department and Center for Astrophysics, Tsinghua University, Beijing 10084, China}

\author{Lizhi Xie}
\affiliation{INAF–Astronomical Observatory of Trieste, via G.B. Tiepolo 11, I-34143 Trieste, Italy}

\author{Xiaoyu Wang}
\affiliation{Key Laboratory for Research in Galaxies and Cosmology, Department of Astronomy, University of Science and Technology of China, Hefei, Anhui 230026, China}
\affiliation{School of Astronomy and Space Science, University of Science and Technology of China, Hefei 230026, China}

\author{Andrea Lapi}
\affiliation{SISSA, Via Bonomea 265, I-34136 Trieste, Italy}
\affiliation{INFN-Sezione di Trie
ste, via Valerio 2, 34127 Trieste, Italy}
\affiliation{INAF-Osservatorio Astronomico di Trieste, via Tiepolo 11, 34131 Trieste, Italy}

\author{Ravi K. Sheth}
\affiliation{Center for Particle Cosmology, University of Pennsylvania, 209 S. 33rd St., Philadelphia, PA 19104, USA}

\correspondingauthor{Jingjing Shi \& Huiyuan Wang}
\email{jingshi@mail.ustc.edu.cn; whywang@ustc.edu.cn}

\begin{abstract}
We use a 200 $\mpc$ a side N-body simulation to study the mass accretion history (MAH) of
dark matter halos to be accreted by larger halos, which we call infall halos. We define a quantity 
$\zn\equiv (1+\zf)/(1+\zpeak)$ to characterize the MAH of infall halos, where $\zpeak$ and 
$\zf$ are the accretion and formation redshifts, respectively. We find that, at given $\zpeak$, 
their MAH is bimodal. Infall halos are dominated by a young population at high redshift and by an 
old population at low redshift. For the young population, the $\zn$ distribution is narrow and 
peaks at about $1.2$, independent of $\zpeak$, while for the old population, the peak position and 
width of the $\zn$ distribution both increases with decreasing $\zpeak$ and are both 
larger than those of the young population. This bimodal distribution is found to be closely 
connected to the two phases in the MAHs of halos. While members of the young population are 
still in the fast accretion phase at $\zpeak$, those of the old population have already entered the slow 
accretion phase at $\zpeak$. This bimodal distribution is not found for the whole halo population, 
nor is it seen in halo merger trees generated with the extended Press-Schechter formalism.
The infall halo population at $\zpeak$ are, on average, younger than the whole halo population 
of similar masses identified at the same redshift. We discuss the implications of our findings in 
connection to the bimodal color distribution of observed galaxies and to the link between central 
and satellite galaxies.
\end{abstract}


\section{Introduction}\label{sec:intro}

In the standard cold dark matter paradigm, dark matter halos are the basic units
of the large-scale structures of the Universe and the hosts within which galaxies form.
A lot of effort has been devoted to understanding the assembly histories of individual
halos, because these histories are expected to be linked directly to the properties
of galaxies that form in halos (see \citealt{Mo2010} for a review).
Numerical simulations and analytical models have both demonstrated that
dark matter halos grow hierarchically via the accretion and merger of smaller halos
\citep{Lacey1993, Springel2005nature}. The mass accretion histories (MAHs) of dark matter
halos are complex, and a number of formation times have been proposed to characterize
the properties of the formation history of a halo (see e.g. \citealt{Li2008}).
The distributions of these formation times are usually single-peaked \citep{Lin2003},
which is very different from the distribution of galaxies, which exhibits a bimodal distribution in star
formation rate (SFR) and color \citep{Blanton2003, Blanton2005, Baldry2004, Wyder2007,
Wetzel2012}. Moreover, numerous studies
have shown that the halo assembly history is correlated with many other halo properties,
such as halo mass, halo structure, dynamical state
\citep{Lacey1993, Jing2002, Gao2004, Allgood2006, Hahn2007, Wang2011, Shi2015},
as well as  large scale environment \citep{Sheth2004, Gao2005, Wang2007, Gao2007, Jing2007}.

So far the investigations have focussed on the main trunk of the halo merger trees
that reflect the evolution history of the main progenitors of a halo. To fully understand the
halo assembly, halos on the subbranches of the merger trees should also been taken into account.
Whereas the main trunk is related to the formation of the central galaxy in a halo,
the sub-branches are related to the formation of satellite galaxies.
The evolutionary history of subbranch halos can be divided into two phases:
one before a halo has merged into a bigger halo and the other after the halo has merged
to become a subhalo. In the first phase, the halos  (hereafter referred to as infall halos)
accrete material similarly to the main trunk halos, as they themselves are independent
halos. They become substructures of larger halos, usually referred to as subhalos, only
after the accretion. During the subhalo phase, the subbranch halos evolve in various
ways due to interactions with the host halo, such as dynamical friction
\citep{Chandrasekhar1943, Oguri2004, Hashimoto2003, Jiang2008},
tidal heating and stripping \citep{Hayashi2003, Taylor2004, Gan2010, Han2016},
back-splashing \citep{Ludlow2009}, and impulsive encounters \citep{vandenBosch2017}.
There have been many studies of subhalo properties
\citep{Tormen1997, Vitvitska2002, Gao2004, Benson2005, Wang2005,
Giocoli2008, Wetzel2011, Jiang2015, Shi2015, Xie2015}.
In contrast, the mass assembly histories of the infall population, have so far
drawn only little attention; the only related work known to us is
\cite{Sheth2003} where a simple Poisson model is developed to investigate the formation
time distribution of infall halos.

As mentioned above, infall halos are the hosts in which satellite galaxies we observed today
form and evolve. Once these galaxies become satellite galaxies, they are expected to
experience satellite-specific processes, such as tidal and ram pressure stripping
\citep{Gunn1972, Abadi1999, Quilis2000}. To quantify the efficiency of these processes,
one usually compares these satellites with central galaxies of similar stellar
mass \citep{vandenBosch2008, Weinmann2009, Pasquali2010, Wetzel2012, Peng2012,
Knobel2013, Bluck2014, Fossati2017}. This is only valid under the assumption
that the halo assembly histories of infall halos, which are the progenitors of the
subhalos hosting satellite galaxies, are similar to the assembly histories of the
halos hosting central galaxies. Clearly, this assumption
needs to be checked by comparing the formation histories between the two kinds of halos.

There is growing evidence that the star formation in a galaxy is correlated with
the assembly history of its host halo. For example, \cite{Bray2016} found that
galaxy color is correlated with halo formation time in the cosmological
hydrodynamic simulation ``Illustris" (e.g. \citealt{Vogelsberger2014}), in the sense
that redder galaxies tend to live in older halos. \cite{Wang2017} found that the quenching
probability, defined as the probability for a galaxy to be quenched, is related to the formation
time of the host halo identified in constrained simulations. In particular, the subhalo abundance
matching model, which links the formation time of the host halo to galaxy color
\citep{Hearin2013, Hearin2014, ChavesMontero2016, Paranjape2015}, can reproduce
the observed clustering and cosmic shear signals separately for red and blue galaxies,
as well as the observed  ``galactic conformity" \citep{Hearin2014, Watson2015, Hearin2016}.
All these suggest that it is important to study the assembly histories of infall in order to
understand the evolution of satellite galaxies.

In this paper, we present a detailed analysis of the mass accretion histories of infall
halos. In Section \ref{sec:sim}, we describe the
numerical simulation we use, the construction of halo merger trees, and the method
to identify infall halos. In Section \ref{sec:results}, we present our main results of the
formation time distribution for infall halos, including the finding of bimodality in the
distribution. In Section \ref{sec:explaination}, we try to explain the bimodality using
the fact that halos have two distinct accretion phases, and compare the properties of
infall halo population with the general halo population.  Finally, in
Section \ref{sec:sum} we summarize our main results and discuss their implications
for galaxy formation.

\section{Numerical Simulation and Dark Matter Halos}
\label{sec:sim}
\subsection{The Simulation and Halo Merger Trees}

The N-body simulation used here was carried out with Gadget-2 \citep{Springel2005},
adopting a flat $\Lambda$CDM cosmology with parameters consistent with WMAP9 data
\citep{Hinshaw2013}: $\Omega_{\rm \Lambda,0}=0.718$, $\Omega_{\rm m,0}=0.282$, $\Omega_{\rm b,0}=0.046$,
$h=H_{0}/(100\kms \Mpc)=0.697$, $\sigma_{8}=0.817$, and $n_{s}=0.96$. The CDM density field is traced
by $2048^{3}$ particles, each with a mass of $\mp\approx 7.29\times10^{7}\msun$, in a cubic box of $200\mpc$
in a side. The gravitational force is softened isotropically on a co-moving length scale of $2\kpc$
(Plummer equivalent). Outputs are made at 100 snapshots from $z=20$ to $z=0$ equally spaced in the
logarithm of the expansion factor.

Dark matter halos are first identified using the FOF algorithm \citep{Davis1985} with a linking
length of $0.2b$, where $b$ is the mean inter particles separation, and all halos with at least 20
particles are selected. We then use the SUBFIND algorithm \citep{Springel2001} to identify gravitationally bound substructures within each FOF halo. The most massive substructure in a FOF halo is called
the main halo, while all the other substructures are referred to as subhalos.
The virial mass ($\Mvir$) of a main halo is defined as the mass contained in a spherical
volume, centered on the minimum of the gravitational potential well, within which the
average density is $200 \rho_{crit}$, with $\rho_{crit}$ being the critical density of the universe.
We do not measure the current masses of subhalos,  as they are usually strongly stripped;
rather, we measure their masses before they are incorporated into their hosts (see below).

The construction of halo merger trees is based on the SUBFIND catalogs using the algorithm described in \cite{Springel2005nature} (see \citealt{BoylanKolchin2009} for a more detailed description).
Briefly, the member particles of a gravitationally bound substructure are assigned a weight that decreases
with the binding energy calculated using all the particles in it.  A search is then made
in the subsequent snapshot for substructures (including both main halos and subhalos)
that contain some of the particles of the substructure in question. The one which contains the 
largest weighted number of particles of the
substructure is chosen as the descendent of the substructure 
in the snapshot. This method allows one to accurately trace
the complex history of a substructure even in the cases where a main halo becomes a subhalo
or where a subhalo is ejected so as to become a main halo again.
For a FOF halo at redshift zero, the merger tree usually contains many branches, including
one main trunk and many sub-branches, with the former tracing the main progenitors
of the main halo in the FOF halo back in time, where the main progenitor is defined to be the most massive progenitor in the previous snapshot. The other parts of the merger tree are referred to as subbranches. Note that the branch that ends up in a subhalo at redshift zero is also referred to as a subbranch.

\subsection{Infall Halos, Accretion Time and Formation Time}

We define an infall halo to be one that is about to merge with a bigger halo. 
More specifically, infall halos are main halos on the sub-branches of merger trees. They 
eventually either become subhalos of larger FOF halos (the survived subhalos) or get totally disrupted 
at $z=0$. We further require that the descendant an infall halo has crossed the virial radius of 
the main trunk halo at least once. This selection criterion can effectively remove halos that are 
temporarily linked to the main trunk halos (see e.g. \citealt{Tinker2008}), 
thus treated as sub-branch halos, due to numerical effects. However, it will also remove some true 
infall halos. Our tests show that the bimodal distribution and our conclusions won't change significantly if we discard this criterion.

An infall halo can be accreted to become the subhalo of a larger halo either on a sub-branch 
or on the main trunk
\citep{Lacey1993, Springel2005nature}. It is  referred to as
a first-order infall halo in the former case, and a higher-order infall halo in the latter.
Since higher-order infall halos may become subhalos in a subhalo and may contribute
significantly to the substructures in the final FOF halos, they are also included in our
analysis.  Moreover, some halos may have been accreted at an earlier time by larger halos,
but subsequently ejected and become independent main halos at $z=0$
\citep{Lin2003, Gill2005, Wang2009, Ludlow2009, Li2013}. These halos, called ``backsplash halos", 
are excluded from our infall halo samples. However, these halos are treated as main trunk halos 
and their own infall halos are included in our analysis. We will also include another population of 
infall halos, called ``wavering" infall halos, which entered their hosts at high $z$, but 
left and re-entered the hosts later.

Once an infall halo is identified, we trace its merger tree and find the redshift
at which its $\Mvir$ reaches the peak value in its lifetime ($\Mpeak$). This redshift is
denoted $\zpeak$ in the following. In our analysis, we ignore the part of the history
after the infall halo is accreted to become a subhalo.  Since a halo in general
loses mass in the subhalo phase, ignoring this part of the history has no significant impact on
the measurements of $\zpeak$ and $\Mpeak$. By construction, all infall halos at $\zpeak$ are themselves main halos.
Following \cite{Xie2015}, we refer to $\zpeak$ as the accretion time
of the infall halo.  Other definitions of the accretion time include the time
when the merger of the infall halo with its host occurs, i.e. the time
when an infall halo first becomes a subhalo (see \citealt{Li2009, Giocoli2008}), and
the time at which the maximum circular velocity ($\Vmax$) of a halo reaches the
maximum value in its lifetime \citep{Conroy2006, Nagai2005}. We will denote 
the corresponding redshifts by $z_{\rm inf}$ and $z_{\rm vp}$, respectively.  
We will discuss how our results change with the definitions of accretion time in Section \ref{sec:at}.

For each infall halo, we estimate a formation time, corresponding to a redshift $\zf$, 
defined as the time at which the halo reaches half of the halo mass at accretion time for the first time.
In most part of this paper, we adopt $\zpeak$ to define accretion time, so the halo mass at $\zf$ equals to $\Mpeak/2$.
Nevertheless, in Section \ref{sec:at}, we adopt two other definitions for the accretion time, 
$z_{\rm inf}$ and $z_{\rm vp}$, so that the corresponding formation times are also different. 
Thus, for each infall halo, we are able to obtain two characteristic times: the accretion time 
(which is also called the infall time) and the formation time. By definition, the formation time is 
always larger than the accretion time.

\subsection{Halo Selection}

\begin{figure}
\centering
\includegraphics[width=1.\linewidth]{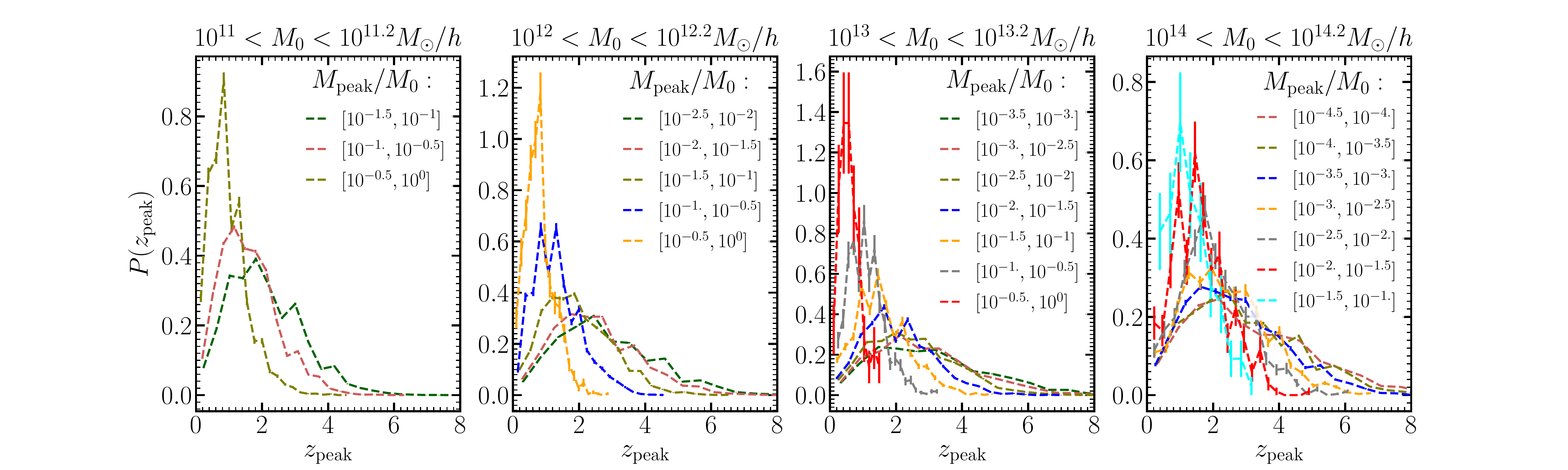}
\includegraphics[width=1.\linewidth]{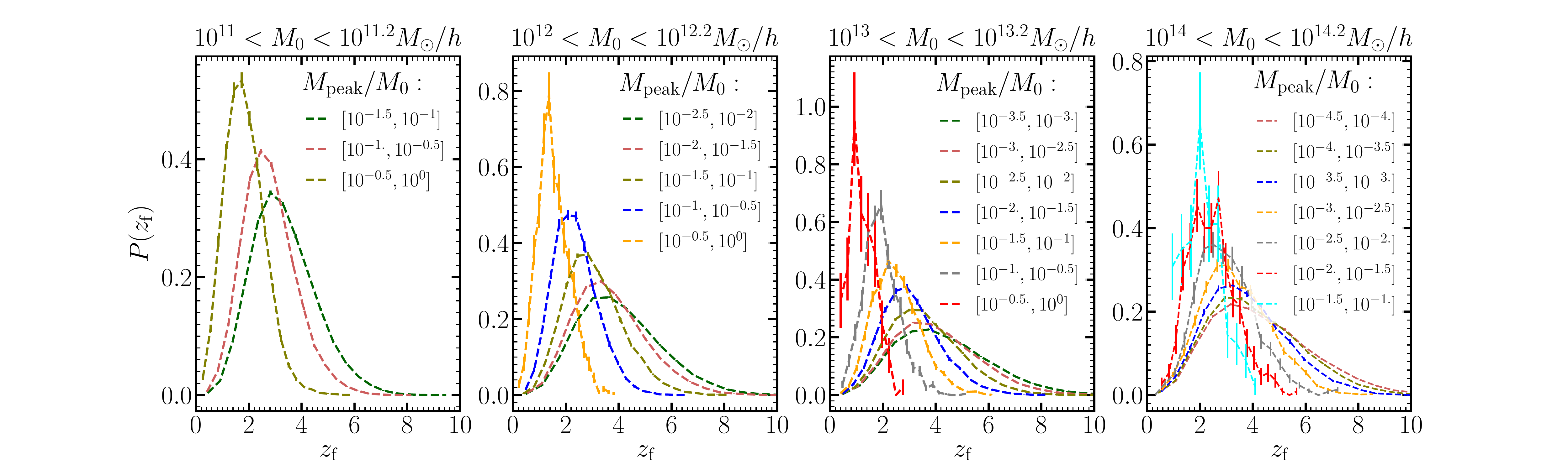}
\caption{Upper panel: the distribution of the accretion redshfit $\zpeak$ for infall halos of various 
$\M0$ and $\Mpeak$ . Lower panel: the same as the upper panel, but for the formation 
redshift $\zf$ of infall halos. Error bars show Poisson errors.}
\label{fig:Pzinf_Pzf_distribution}
\end{figure}

In order to obtain $\zf$ reliably, we only consider infall halos with $\Mpeak>100\mp$. Throughout the paper, we use $\M0$ to denote the mass of the main trunk halo at $z=0$, and restrict our analysis to halos with $10^{11}<M_{0}<6\times 10^{14}\msun$, which contain at least 1,300 particles. With these selections, our final sample 
contains 191,166 main trunk halos at $z=0$ and 2,402,610 infall halos. We sometimes also use `host halo' to 
refer to the halo into which an infall halo is accreted. It should emphasized, however, that host halos are not 
necessarily main trunk halos, since the main halos on subbranches can also accrete halos and 
thus be called host halos.
Figure~\ref{fig:Pzinf_Pzf_distribution} shows separately the probability distributions of 
$\zpeak$ and $\zf$ for four $\M0$ bins (as shown in each column) and several $\Mpeak$ bins (as 
indicated by the labels in each panel). Both quantities have single-peaked (unimodal) 
distributions. In general, more massive infall halos tend to have lower $\zpeak$ 
and lower $\zf$.

\section{Bimodal Formation Time Distribution}
\label{sec:results}

\subsection{Bimodal distributions of formation times for infall halos}

\begin{figure}
\centering
\includegraphics[width=1.\linewidth]{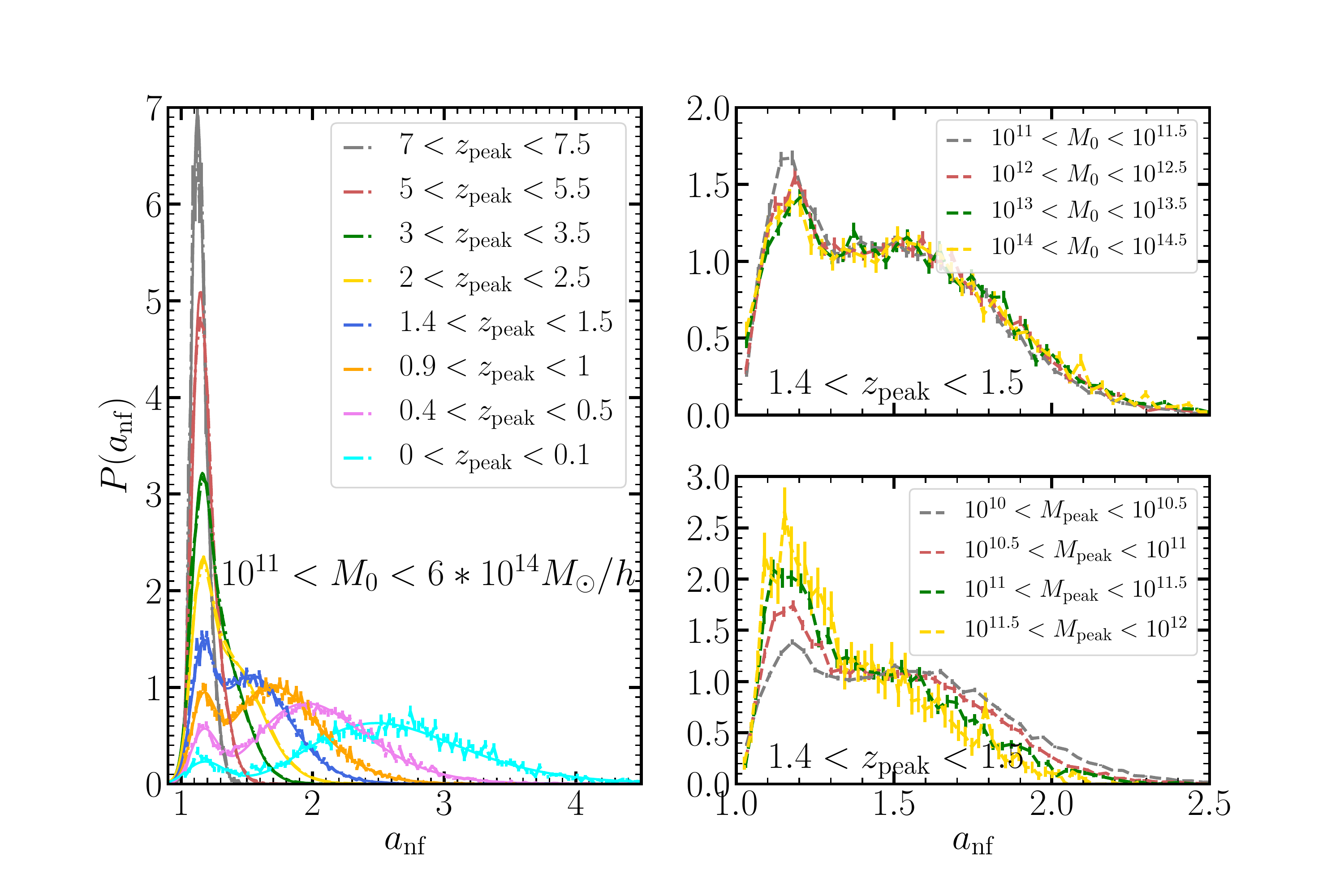}
\caption{Left panel: The distributions of $\zn\equiv (1+\zf)/(1+\zpeak)$ for infall halos
with $\Mpeak>100\mp$ at various $\zpeak$. Right Panel: the $\zn$ distributions
for halos of various $\M0$ (upper panel), and various $\Mpeak$ (lower panel).
In both panels, results are shown only for $1.4<\zpeak<1.5$. Solid lines are
double log-normal fitting curves. Error bars show Poisson errors.}
\label{fig:bimodal_distribution}
\end{figure}

We show the formation redshift distributions in the left panel of
Figure~\ref{fig:bimodal_distribution}
for infall halos with $\Mpeak>100\mp$ and $10^{11}<M_{0}<6\times 10^{14}\msun$ at
various $\zpeak$. To compare results for different $\zpeak$ and to better 
understand the results, we actually show the distributions of
\begin{equation}
\zn\equiv \frac{1+\zf}{1+\zpeak}.
\end{equation}
$\zn$ can be used to describe the relative age of infall halos that are accreted at 
the same accretion time. Interestingly, the $\zn$ distribution is bimodal, while  
the formation time distributions shown in Figure~\ref{fig:Pzinf_Pzf_distribution}
are clearly unimodal. This indicates that the bimodality appears only for fixed 
$\zpeak$. 

For halos with $\zpeak > 5$, the distributions of $\zn$ are narrow and peak at 
$\zn \sim 1.2$. This implies a uniform
accretion pattern, as we will discuss below. As $\zpeak$ decreases, a second
peak appears in the distribution and becomes increasingly more important.
At $\zpeak \sim 2$, the two components become comparable in height.
As $\zpeak$ approaches zero, the peak at low $\zn$ almost disappears.
For convenience,  we refer halos in the low $\zn$ mode as the
{\it young population}, and those in the high $\zn$ mode as the {\it old population}.

To describe the redshift evolution of the two populations,
we fit the $\zn$ distributions with a double log-normal function:
\begin{equation}
P(\zn | \zpeak)=\omega
\mathcal{N}_{\rm log}(\mu_1,\sigma_1^2)
+(1-\omega) \mathcal{N}_{\rm log}(\mu_2,\sigma_2^2)\,\,,
\label{eq:dlognorm}
\end{equation}
where $\mu_1<\mu_2$ and $\mathcal{N}_{\rm log}(\mu,\sigma^2)$ represents the
log-normal function. The fitting is performed with the non-linear least square method.
As shown in Figure~\ref{fig:bimodal_distribution},
the distributions can be well fitted by the double log-normal model.
The two populations clearly evolve differently. Our best fitting result suggests that,
for the young population, the peak position and width of the distribution change
little with $\zpeak$, with the peak staying at $\zn\sim 1.2$. In contrast, both the
peak position and dispersion decrease significantly with increasing
$\zpeak$ for the old population.

\subsection{Dependence on $\Mpeak$, $\M0$, and sub-classes of infall halos}

For the main trunk halos, it is known that the formation time depends on halo mass. We thus check whether
the bimodality is related to halo mass. Since the bimodal feature is the most prominent for $1<\zpeak<2$, we
show the results for infall halos with $1.4 < \zpeak < 1.5$. The upper and lower right panels
of Figure~\ref{fig:bimodal_distribution} show the $(1+\zf)/(1+\zpeak)$ distributions for
various $\M0$ and $\Mpeak$, respectively. The bimodality is
clearly present in all the mass bins shown. The dependence on $\M0$ is rather weak;
there is a very weak trend for the fraction of the young population to decrease
with increasing $\M0$. The dependence on $\Mpeak$ is stronger, with
the young population fraction increasing significantly with $\Mpeak$.
Similar to the result for the total sample shown in the left panel of
Figure~\ref{fig:bimodal_distribution},  there is no significant change in the
peak position and width for the young population. Both the peak and width of the
distribution increase with decreasing $\Mpeak$ for the old population, which is similar
to the main trunk halos (e.g. \citealt{Wang2007}).

\begin{figure}
\centering
\includegraphics[width=1.\linewidth]{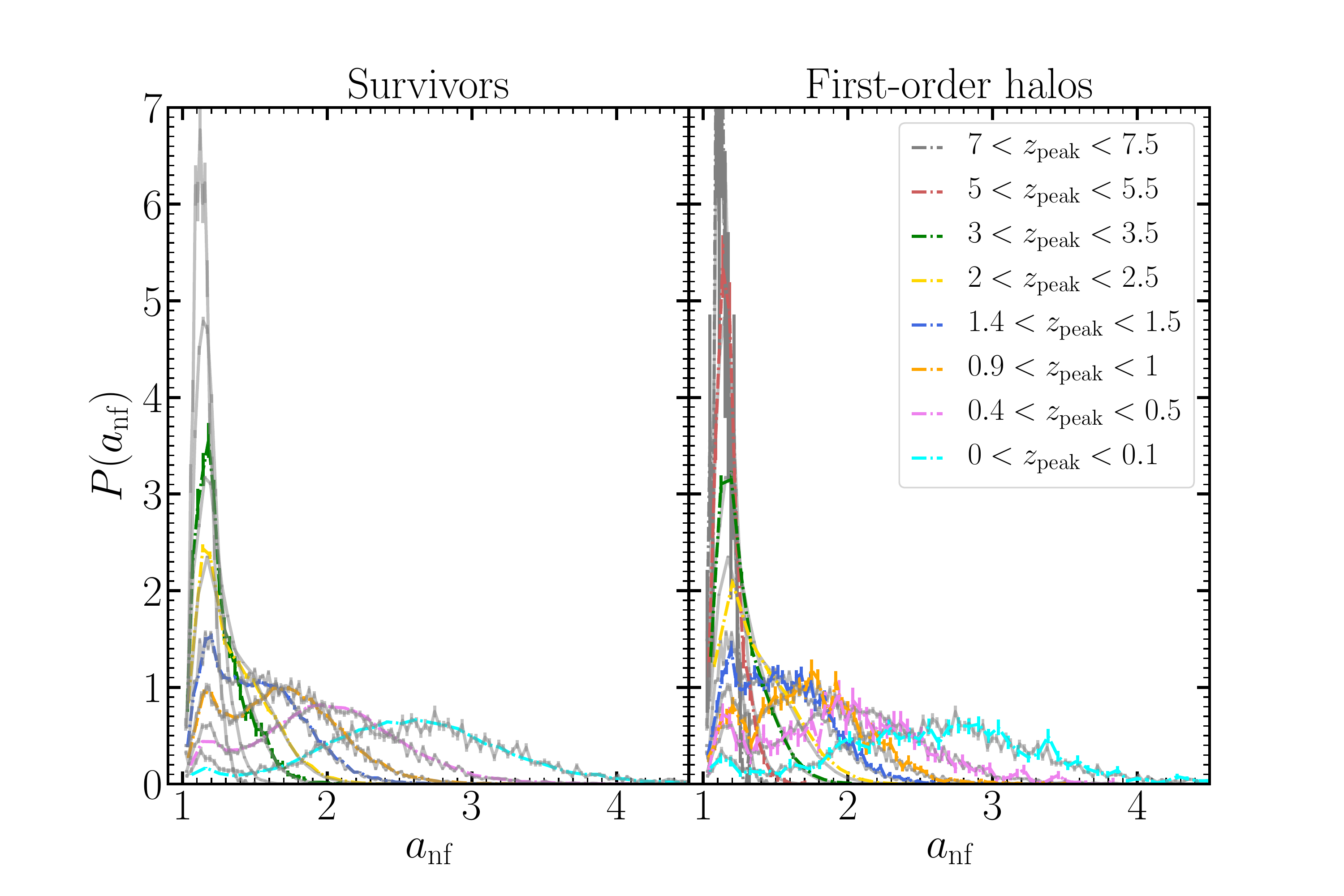}
\caption{Left panel shows the distributions of $\zn$ for infall halos that survive
as subhalos at $z=0$. Right panel shows results for halos that infall directly onto the
main trunk. The solid gray lines are the results for the whole population as shown in
the left panel of figure \ref{fig:bimodal_distribution}. The error bars show Poisson errors.}
\label{fig:bimodal_distribution_surv}
\end{figure}

An interesting question  is whether the existence of the bimodality depends on the final
states of the infall halos. In the left panel of Figure~\ref{fig:bimodal_distribution_surv},
we show the results for infall halos that survive as subhalos at redshift zero. Since only
few infall halos with $\zpeak>4$ can survive in their final hosts, the results for the
two highest redshift bins are not shown. As one can see, the bimodality in the distribution
of $\zn$ is very similar to that for the total population, indicating that the bimodal feature
is independent of the final states of the infall halos.

Next, we examine first-order infall halos, namely halos that fall directly into 
the main trunk halos, no matter whether they survive or are fully disrupted 
(see figure 1 of \citet{Jiang2014} for a depiction of such halos). 
The results are shown in the right panel of
Figure~\ref{fig:bimodal_distribution_surv}. Again, the difference from the total population
is quite small.  We have also checked other two sub-populations: the ``wavering" population,
which entered their host at high redshift, but then left and re-entered the hosts later;
infall halos that are accreted by the `backsplash' halos. These two
sub-populations also exhibit clear bimodality in their $\zn$ distributions.

\subsection{Resolution tests}
To examine whether or not the observed bimodality can be significantly biased or even
caused by the mass resolution of the simulation, we have made the same analysis using a
simulation in a $500\mpc$ box run with $3072^3$ particles \citep{Li2016}. This simulation
uses the same cosmology as the simulation used above,  but has a mass resolution that is
lower by a factor of $4.63$.  For the same ranges of $\M0$, $\Mpeak$, and $\zpeak$, the
$\zn$ distributions obtained from the two simulations are indistinguishable.
We have also applied our analysis to two suites of high-resolution simulations of
individual dark matter halos from the Phoenix and the Aquarius projects
\citep{Springel2008, Gao2012}.  These have mass resolutions that are higher than
our simulation by a factor of more than 10 and 1000, respectively.  Similar bimodal
distributions are also found for these individual halos. This demonstrates that our
results are not affected by the resolution of our simulation.

\subsection{Dependence on the definition of accretion times}
\label{sec:at}

We have also checked how our results may vary with different definitions for the accretion
time. In addition to $\zpeak$, there are two other commonly used definitions: the redshift,
$z_{\rm vp}$, at which the maximum circular velocity ($\Vmax$) of a halo reaches the
maximum value in its lifetime \citep{Conroy2006, Nagai2005}; and the redshift $z_{\rm inf}$
at which a halo becomes a subhalo for the first time \citep{Li2009, Shi2015}. 
Similarly, we define the formation redshift as the redshift
at which the infall halo first reaches half of its mass at the accretion redshift for these two 
definitions of accretion time.
We find that using $z_{\rm vp}$ instead of $\zpeak$ does not change the bimodal
distribution of the formation redshift, but the bimodality becomes less prominent when $z_{\rm inf}$ is used
instead of $z_{peak}$. About 50\% of the infall halos have $z_{\rm inf}=\zpeak$, and
so the reduced prominence comes from the other infall halos.
Among them, about 60\% have  $z_{\rm inf}<\zpeak$, and these halos are
expected to have experienced tidal stripping before falling into their hosts.
For the 40\% that have $z_{\rm inf}>\zpeak$, their masses actually grow
after being accreted by larger halos.  There are two possibilities for this
after-accretion growth. First, the halos may be ejected from their hosts and
grow mass after $z_{\rm inf}$;  second, the halos are not accreted at $z_{\rm inf}$,
but linked with their `hosts' by a temporary bridge (see e.g. \citealt{Tinker2008}).
Because of these uncertainties, we believe that $\zpeak$ is a better choice than
$z_{\rm inf}$ for defining the accretion of a halo.

\section{Fast and Slow Accretion}
\label{sec:explaination}

\begin{figure}
\centering
\includegraphics[width=1.\linewidth]{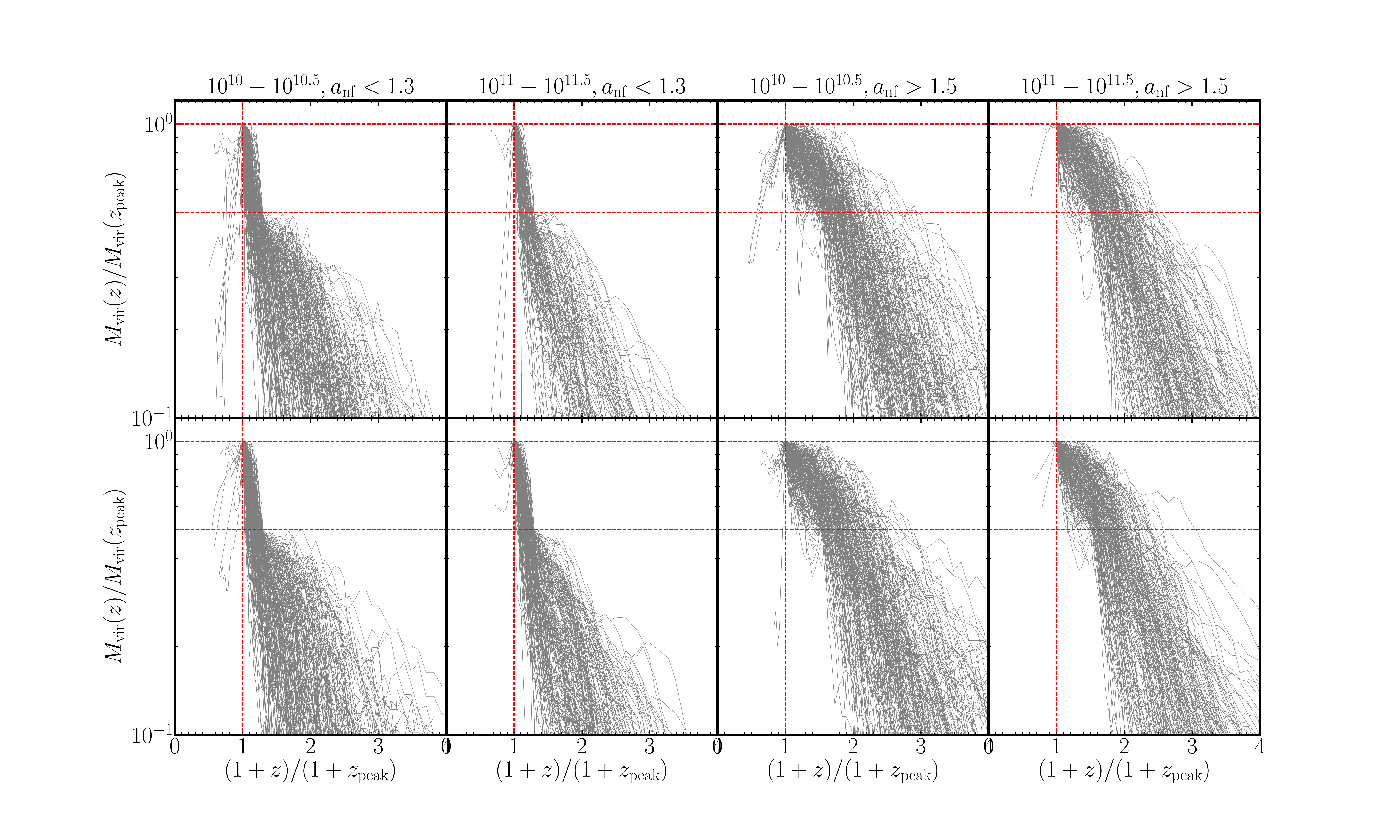}
\caption{Mass accretion histories for young (left two columns) and old (right two columns)
  infall halos with different $\Mpeak$ and $\M0$. Here the young population has halos with
  $\zn<1.3$ while the old halo population has $\zn>1.5$. The upper panels
show the results for $10^{12}<\M0<10^{13}\msun$; the lower panels for
$10^{13}<\M0<10^{14}\msun$. Two mass ranges for infall halos are chosen:
$10^{10}-10^{10.5}\msun$ and $10^{11}-10^{11.5}\msun$, as indicated in each panel.
The upper and lower horizontal red lines indicate  $M(z)=\M0$ and $M(z)=\M0/2$, respectively.}
\label{fig:mz}
\end{figure}

The formation time, $\zf$, is only one of many parameters that characterize the
mass assembly history (MAH) of a halo. To obtain more insight about halo growth,
we show in Figure \ref{fig:mz} the MAH for infall halos randomly selected from the
two populations. For each halo selected, we plot $\Mvir(z)$ (normalized by $M(\zpeak)$)
as a function of $(1+z)/(1+\zpeak)$, where $\Mvir(z)$ is mass of the most massive progenitor at $z$.
Motivated by the presence of the bimodal feature, we use infall halos with $\zn<1.3$
to represent the young population and the ones with $\zn>1.5$ to represent the
old population. Clearly, the two populations have quite different MAH. The old population
is characterized by a fast mass growth at high redshift, and the mass accretion rate slows
down before reaching $\Mpeak$. In contrast, the young population shows a
much faster growth than the old one, in particularly at low $\zn$.
These demonstrate clearly that the presence of the bimodal distribution in
$\zn$ is closely related to the MAH of halos.

The mass accretion history of infall halos plays a key role in our understanding
of the formation and evolution of satellite galaxies. It is thus important to investigate
the origin of the bimodal distribution. \cite{Zhao2003a,Zhao2003b} found that the MAH of a
halo can be divided into a fast accretion phase and a slow accretion phase, and the average
MAH can be described in the form,
\beq
\frac{\Mvir(z)}{\Mvir(\ztp)}=\frac{t^{0.3}}{1-b+bt^{-1.8b}}\,\,,
\label{eq:twophase}
\eeq
where $t\equiv \rho_{\rm vir}(\ztp)/\rho_{\rm vir}(z)$ and $\ztp$ stands for
the transition redshift between the two phases. The parameter $b$ is $0.75$ for the
fast accretion phase ($z>\ztp$) and $0.42$ for the slow accretion phase ($z<\ztp$).
Since we assume $\rho_{\rm vir}(z)=200\rho_{\rm crit}(z)$, the dimensionless time
variable can be written as $t=\left[H^2(\ztp)/H^2(z)\right]$.
This formula suggests that, at high redshift, most halos are in the fast accretion phase,
but at redshift zero, most halos are in the slow accretion phase.

\begin{figure}
\centering
\includegraphics[width=0.8\linewidth]{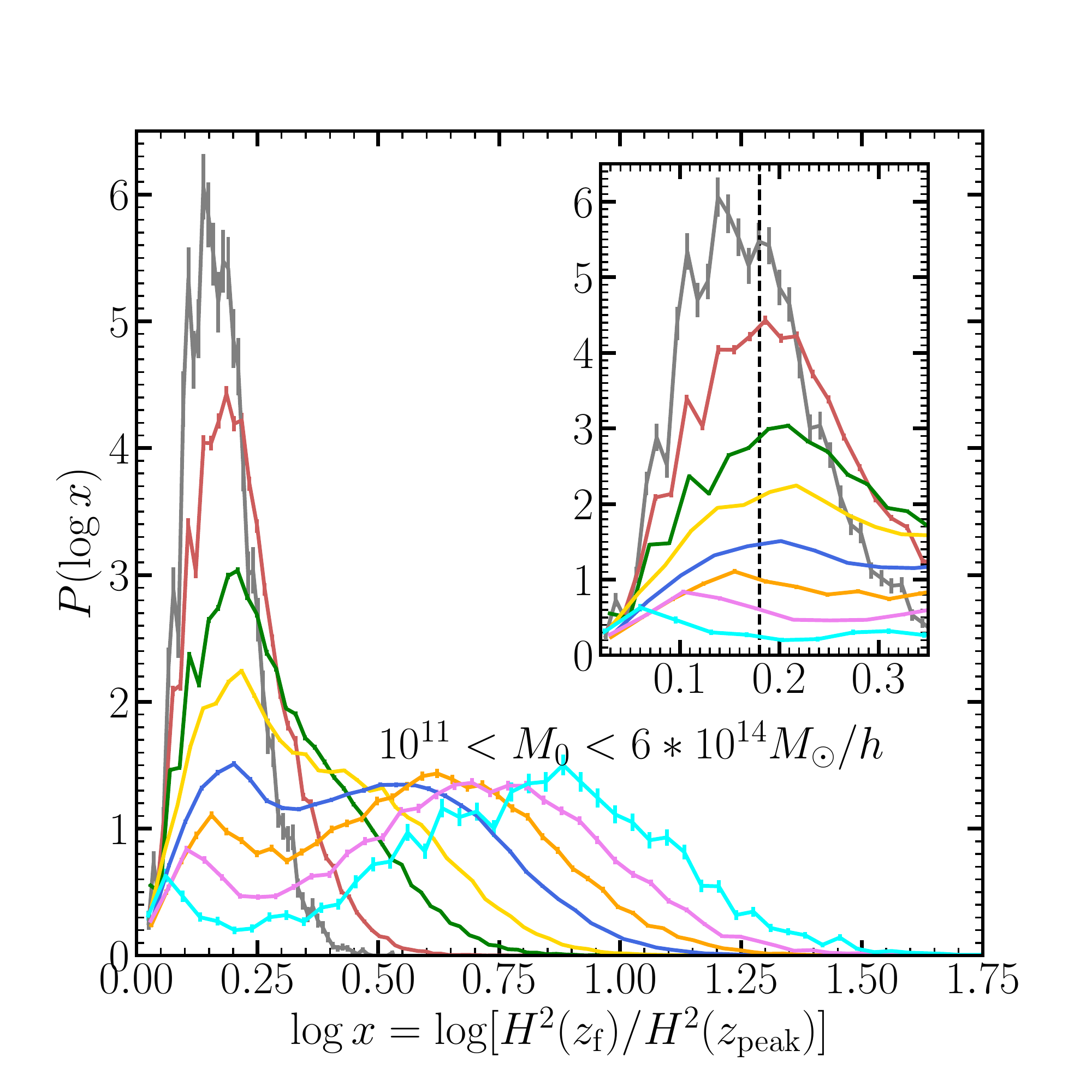}
\caption{Similar to the left panel in Figure \ref{fig:bimodal_distribution}, except
that now it shows the distribution of ${\rm log}x={\rm log}\left[H^2(\zf)/H^2(\zpeak)\right]$ for
various $\zpeak$. The insert shows the ${\rm log}x$ distribution around the young population.
The dashed vertical line indicates ${\rm log}x=0.18$. See text for the detail.}
\label{fig:Hz}
\end{figure}

To understand the connection of the bimodal distribution to the two-phase accretion,
we consider two extreme cases. First, suppose that an infall halo has $\zpeak\gg\ztp$,
and so $\zf\gg\ztp$. We then have
$t_{\rm f}=H^2(\ztp)/H^2(\zf)\ll1$, and $t_{\rm peak}=H^2(\ztp)/H^2(\zpeak)\ll1$.
Since this halo stays in the fast accretion phase before being accreted at $z>\zpeak$,
we have $b=0.75$ and $\Mvir(z)/\Mvir(\ztp)\simeq t^{1.65}/0.75$. We can define a new quantity,
\beq
\log x\equiv\log\left[H^2(\zf)/H^2(\zpeak)\right]=\log\left[t_{\rm peak}/t_{\rm f}\right]\,.
\eeq
For this halo, this equation can be rewritten as $\log x\simeq\log\left[\Mvir(\zpeak)/\Mvir(\zf)\right]/1.65$.
By definition, $\Mvir(\zf)=\Mvir(\zpeak)/2=\Mpeak/2$, so we have $\log x\simeq 0.18$.
In the second case, where the infall halo is assumed to have $\zpeak\ll\ztp$ and $\zf\ll\ztp$,
we have $t_{\rm f}\gg1$ and $t_{\rm peak}\gg1$, and so the halo is in the slow accretion
phase ($b=0.42$). Its MAH can then be simplified as $\Mvir(z)/\Mvir(\ztp)\simeq t^{0.3}/0.58$
and we have $\log x\simeq 1$. For each infall halo, we have its $\zf$ and $\zpeak$,
and so can derive its $\log x= \log\left[H^2(\zf)/H^2(\zpeak)\right]$.
Figure \ref{fig:Hz} shows the distributions of ${\rm log}x={\rm log}\left[H^2(\zf)/H^2(\zpeak)\right]$
for halos with $\Mpeak>100\mp$, $10^{11}<M_{0}<6\times 10^{14}\msun$, and various $\zpeak$.
The colors are coded in the same way as in the left panel in Figure \ref{fig:bimodal_distribution}.
The new quantity also shows a clear bimodal distribution. In particular, the distributions for
the young population roughly peak around $\log x=0.18$ for most values of $\zpeak$. This
suggests strongly that the young population halos are in the fast accretion phase and are far
from the turning point (i.e. $\zf>\zpeak\gg\ztp$).  The situation for the old population
looks more complicated. The peak of the distribution increases with decreasing $\zpeak$ and
ranges from $\sim 0.3$ to $\sim 1$. Only for $\zpeak\sim 0$ is the peak of the $\log x$ distribution
close to 1. This indicates that only the old population halos at $z\sim 0$ satisfy
the requirements that both $\zpeak$ and $\zf$ are much less than $\ztp$.  For the old population
at high redshift, it is likely that their $\zf$ are larger than or comparable to $\ztp$.
This means that these old halos in fact spend much or even most of their lifetimes in the fast
accretion phase at $z>\zf$. Finally, we note that the MAH shown in Eq. (\ref{eq:twophase}) is
only a mean relation, and the scatter in it is quite large (see \citealt{Zhao2003b})  This may
partly explain the dispersions of the distribution around the two peaks.

\begin{figure}
\centering
\includegraphics[width=1.\linewidth]{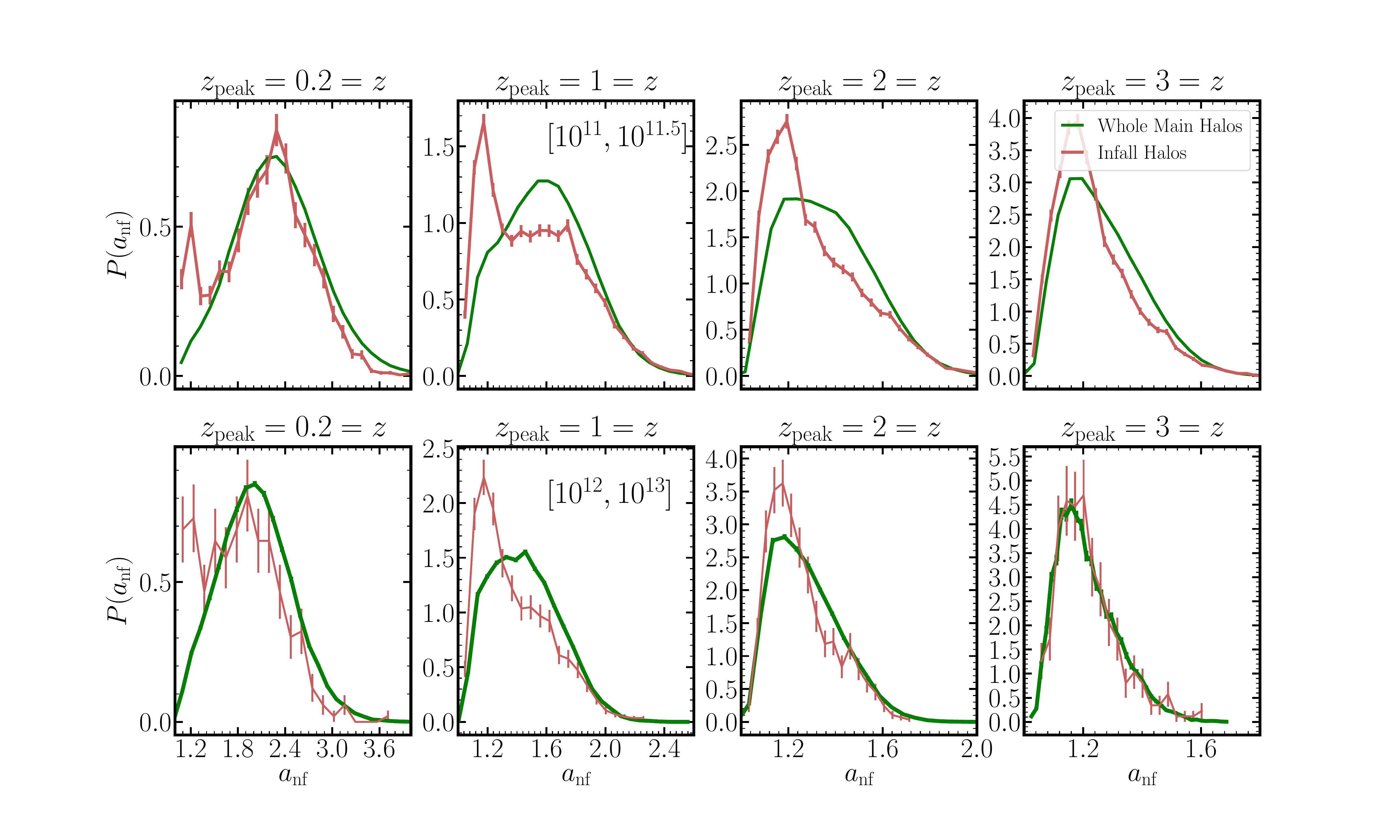}
\caption{Formation time distributions for the whole main halos (green lines) and infall (red lines) halos at four redshifts as indicated in each panel. Note that for infall halos, $\zn \equiv (1+\zf)/(1+\zpeak)$, while for the whole main halos, $\zn \equiv (1+\zf)/(1+z)$, where $\zpeak$ (or $z$) is the redshift at which the infall halos (the main halos) are chosen. The upper panels show the results for small halos with halo masses in the range $[10^{11}, 10^{11.5}]\msun$; the lower panels for larger halos with masses in $[10^{12}, 10^{13}]\msun$. See text for more detail. Error bars show Poisson errors.}
\label{fig:hzf}
\end{figure}

The above analysis suggests that the two populations identified in the
$\zn$ distribution are directly related to the two phases of halo growth. It is thus interesting to see whether such bimodality also exists
for the whole main halo population. Note again that the infall halo sample 
at $z_{\rm peak}$ is a subset of the main halo population at the same redshift.
We select main halos in two halo mass ranges ($10^{11}<\Mvir<10^{11.5}\msun$ and $10^{12}<\Mvir<10^{13}\msun$)
from four snapshots at $z\simeq0.2, 1, 2, 3$. We define the formation time $\zf$ as the redshift, at which a halo reaches half of the mass at $z$ for the first time. Figure \ref{fig:hzf} shows the distribution of $\zn$ for these main halos, where $\zn \equiv (1+\zf)/(1+z)$.
For comparison, we also show in the same figure the results for infall halos with
$\zpeak\simeq 0.2, 1, 2, 3$ and with $\Mpeak$ in the same mass ranges.
Since there is no information about the host halo mass for
common main halos,  we do not limit the host halo masses for infall
halos in the comparison.  In contrast to the infall halos, the $\zn$ distributions
of the whole main halo populations are single-peaked for all the redshift and mass ranges considered.
This is consistent with previous results that the formation time distribution is unimodal (e.g. \citealt{Lacey1993,Lin2003}).

In general, main halos tend to be older than the corresponding infall halos. At low
redshift, the distributions for the main halos are similar to those of the old population of
infall halos. Inspecting the distributions in detail, one can see a small bump at
$\zn\sim1.2$ for main halos. This suggests that the young population
(in the fast accretion phase) does exist in the whole halo sample, though its fraction is much
smaller than that in the infall halos. At high redshift ($z\geq 3$), the distributions for
main halos also peak at $1.2$, similarly to those of infall halos.  This is because most of
the halos at high redshift tend to be in the fast accretion phase (see e.g. \citealt{Zhao2003a}).
Moreover, we can find that the difference in the distribution of $\zn$ between the infall and main
halos reaches its maximum at redshift between 1 and 2. This comparison clearly indicates that the presence
of the bimodality in the formation time distribution is a property of infall halos that will
eventually merge into a bigger halo.  Apparently, the environment determined by the bigger halo
can affect the formation histories of the smaller halos that will merge into it.  We will come
back to this question in a separate paper.

\section{Numerical vs. semi-empirical predictions}
\label{sec:eps}

The above analysis is based on numerical simulations. Halo merger trees
can also be constructed with the extended Press-Schechter formalism
\citep{Bond1991}, and such a formalism has been widely used in studying
halo MAHs and in semi-analytic models of galaxy formation \citep{Kauffmann1993,
Sheth1999, Cole2000, Parkinson2008, Jiang2016, Somerville2015}.
It is, therefore, also interesting to analyze if EPS merger trees have similar
properties as the merger trees obtained from simulations.
To this end, we use the code developed by \cite{Parkinson2008}
\footnote{\url{http://star-www.dur.ac.uk/~cole/merger_trees/}} to generate
merger trees. \cite{Jiang2014} compared several EPS merger tree generating
codes, and found that the algorithm developed by \cite{Parkinson2008}
agrees well with simulations in progenitor mass function, MAH, merger rate per
descendant halo, and the un-evolved subhalo mass function. Here we test its performance 
in describing the MAHs of infall halos.

We generate 2000, 2000, 1000, and 100 merger trees for four host masses,
$10^{11.25}\msun$, $10^{12.25}\msun$, $10^{13.25}\msun$, and $10^{14.25}\msun$, respectively.
Consider a halo at the present time. The tree-generating program draws a set of random progenitor 
halos according to the progenitor halo mass function at a slightly earlier time. The procedure is repeated for each of 
the progenitors as we move back in time until the halo mass resolution limits, $7.29\times 10^{5}\msun$, $7.29\times 10^{6}\msun$, $7.29\times 10^{7}\msun$, and $7.29\times 10^{7}\msun$ are reached correspondingly. 
The collection of all the progenitors at different times and their links form the merger tree of the halo. 
Any halo on the merger tree has only 
one descendant in the next snapshot but may have several progenitors at an earlier snapshot.
The sum of the masses of all the progenitors is equal to the halo mass, and the most massive progenitor 
is called the main progenitor of the halo. Similarly, an infall halo is one on a sub-branch,  and the 
accretion time is defined as the time when the infall halo is not the main progenitor of its descendant. 
Since the mass of a halo always grows with time  in an EPS merger tree, the accretion time of an 
infall halo is exactly the same as the time when the infall halo reaches its maximum mass before 
accretion. The formation time is then defined as the time when the infall halo first reaches half of its 
maximum mass. Different from the merger trees in numerical simulations, the EPS merger trees 
are unable to trace the evolution of subhalos. In this case, there are no backsplash halos to exclude 
and we are not able to investigate whether or not an infall halo has crossed the virial radius of its 
descendant more than once.

\begin{figure}
\centering
\includegraphics[width=1.\linewidth]{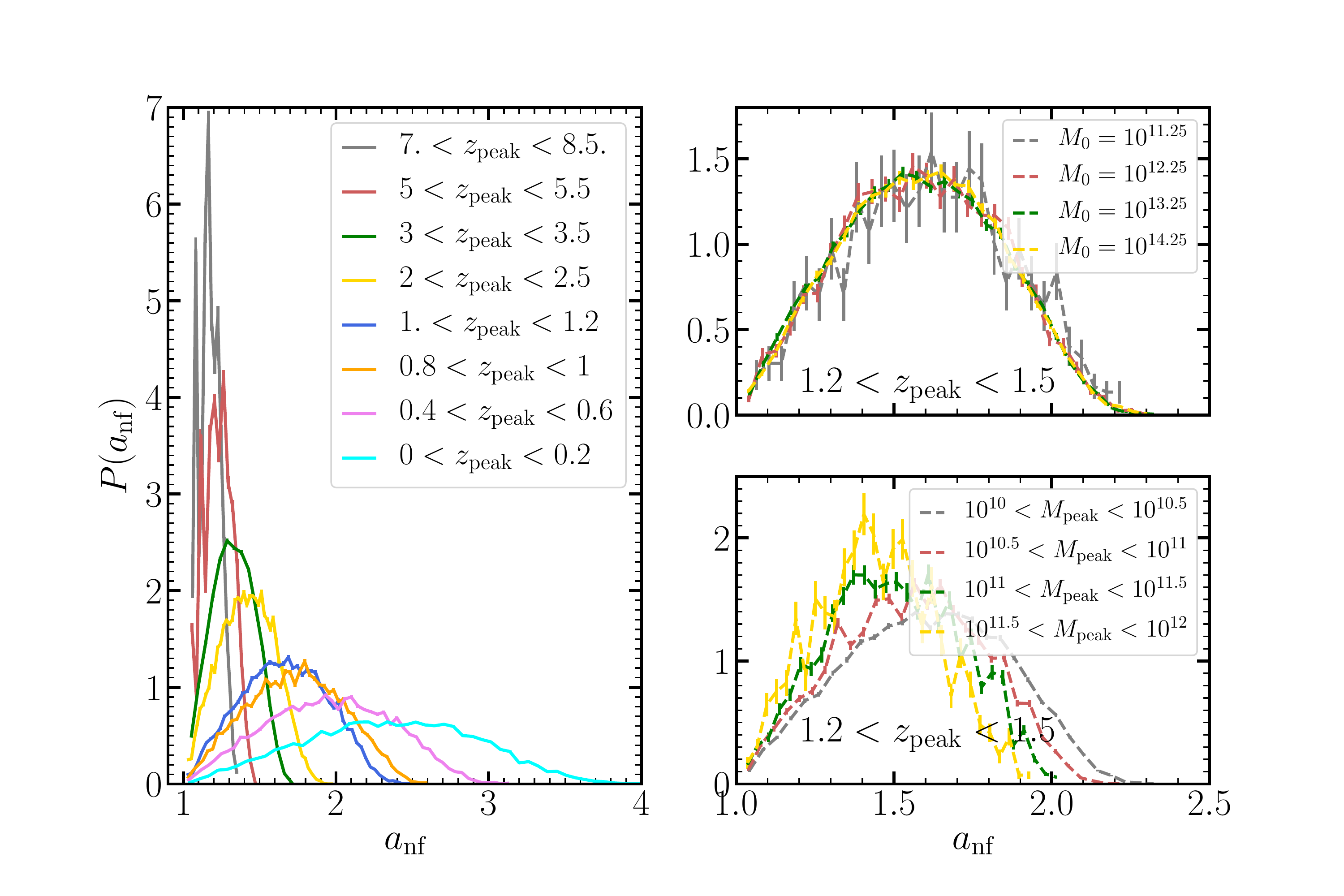}
\caption{The same as Fig.~\ref{fig:bimodal_distribution}, but for merger trees
generated with EPS formalism. Error bars show Poisson errors.}
\label{fig:eps}
\end{figure}

Figure \ref{fig:eps} shows the distributions of $\zn$ in similar $\zpeak$,
$\M0$, and $\Mpeak$ bins as shown in Fig.\,\ref{fig:bimodal_distribution}. A similar $\Mpeak$ threshold, i.e. $\Mpeak>100m_{\rm p}=7.29\times 10^9\msun$, has been applied to the infall halo sample as well.
The general trends with $\zpeak$, $\Mpeak$ and $\M0$ are similar to those
obtained from our simulation, but the bimodality is completely absent
in the EPS merger trees. However, because of the differences in the definition of halos 
and in the treatment of halo accretion between the simulation and the EPS formalism, 
the exact cause of the discrepancy is unclear. One possibility is that the bimodality is the result of
some environmental effects that are present in the simulation but not taken into account by 
the EPS formalism.

\section{Summary and Discussion}
\label{sec:sum}

We investigate the assembly history of infall halos using high-resolution N-body
simulations. These are halos on the branches of the halo merger trees \emph{before}
they are accreted into larger halos.  We define the accretion redshift ($\zpeak$) of an infall halo
as the redshift at which its halo mass reaches the peak value, and the formation redshift ($\zf$) 
as the redshift at which the infall halo reaches half of its peak mass for the first time. 
To compare the infall halos at different accretion time, we define a quantity 
$\zn\equiv (1+\zf)/(1+\zpeak)$ and examine its distribution. We find that, at a 
given accretion time, infall halos have bimodal distributions in $\zn$ and $\zf$. 
The following is a list of our main results:
\begin{itemize}
\item 
According to the $\zn$ distribution, infall halos contain two distinct populations. 
For the first population (young population), the $\zn$ distribution changes only slightly with 
$\zpeak$, peaking at $\zn\sim 1.2$. For the second population (old population), the peak value 
and the width both increase with decreasing $\zpeak$, and are both larger than 
those of the young population.
\item 
The infall halos are dominated by the young population at high redshift, while the old population becomes 
more and more important as the redshift decreases. At $\zpeak \sim 2$, the two populations become 
comparable in number.
\item 
Our analysis shows that the bimodal distribution naturally arises from the two-phase accretion histories 
of dark matter halos. The young population consists of halos that are still in the fast accretion phase 
at the time of accretion, while halos in the old population have already entered slow accretion phase 
at the time of accretion.
\item 
We have also studied the assembly histories of common individual halos without distinguishing whether they will be accreted or not. No significant bimodal feature is found in the distribution 
of their formation redshifts. This indicates that the environments defined by the host halo 
may affect the formation histories of its subhalos even before they are accreted into the host.
\item 
We have also checked the merger trees generated with the EPS formalism and found 
that the infall halos in such merger trees do not show bimodal distribution in formation redshifts. 
This difference between EPS and $N$-body merger trees may be caused 
by the fact that  environmental effects, which are taken into account in the simulation but not in the 
EPS formalism, are important in the formation and evolution of infall halos.  
\end{itemize}

It is well known that galaxies exhibit bimodal distributions in their colors and
star formation rates (SFR; e.g. \citealt{Strateva2001, Blanton2003, Baldry2004, Brinchmann2004}).
In the current scenario of galaxy formation,  galaxy properties are expected to correlate
with the assembly histories of their host halos.
For example, the age-matching model assumes that older halos tend to host
galaxies with older stellar populations \citep{Hearin2013,Hearin2014,Watson2015}.
This simple model successfully reproduces the trends of galaxy color with a variety of
galaxy statistics,  such as galaxy clustering and the galaxy-galaxy lensing signal.
Hydrodynamical simulations \citep{Bray2016} also reveal a correlation between
the assembly histories of galaxies and those of their halos.
In addition, halo spins and concentrations are strongly correlated with halo
formation time \citep{Zhao2003b,Wechsler2002,Wang2011,Hahn2007}. These two halo
parameters are thought to play an important role in shaping the disc size and the surface
density of gas in galaxy disks \citep{Mo1998, Guo2011, Croton2016, Henriques2015},
and thus potentially having effects on galaxy morphology. The bimodal distribution
in the formation time of infall halos we have found may, therefore,  provide insight
into the origin of these two distinct galaxy populations, particularly the origin of
color bimodality found for satellite galaxies \citep{vandenBosch2008}.
We emphasize, however, that \emph{only} infall halos have clearly bimodal formation
time distributions, and that there is considerable scatter between galaxy color/SFR
and the formation time of the host halo \citep[see e.g.][]{Wang2017}.
Clearly, the relation between the bimodality in halo formation and that in galaxy color
merits further study.

Once infall halos are accreted by larger halos, their central galaxies become satellites.
These satellites are expected to undergo some satellite-specific quenching and
morphology-transformation processes. These processes have been constrained by comparing
the properties of satellites to those of central galaxies in control samples
\citep{vandenBosch2008, Weinmann2009, Pasquali2010, Wetzel2012, Peng2012, Knobel2013, Bluck2014, Fossati2017}.
Most of the control samples were made so that the centrals in the control sample
had the same stellar mass and redshift distributions as the satellites;
\cite{Wetzel2013} also accounted for evolution of the centrals.  However, the
assumption underlying all these studies is that, before they are accreted, satellites
are the same as central galaxies of the same stellar mass. This assumption is not
supported by our finding that on average, infall halos are younger than the whole halo population of
the same mass. If galaxy properties such as star formation rate and color, indeed correlate
with the assembly histories of their host halos, then previous studies will underestimate
the efficiency of satellite-specific quenching processes. 
Clearly, well controlled samples of centrals and satellites, matched not only 
in stellar mass (and/or halo mass), but also in the mass assembly history, are needed  
(see \citet{Mistani2016}). Our results should, therefore, be useful in interpreting the 
observational data in terms of satellite quenching processes.

\section*{Acknowledgements}

We thanks the anonymous referee for a useful report that significantly improve the presentation of this paper.
J. Shi  thanks J. Tinker for helpful discussion, and the hospitality of Nordita
during the program of ``Advances in Theoretical Cosmology in Light of Data ".
She also thanks UMass and UPenn for their hospitality in late 2016.
RKS thanks the ICTP for its hospitality during summer 2017.
The Phoenix and Aquarius data were provided by the Virgo Consortium.
We thank the GALFORM team for making the EPS merger tree generating
code publicly available.
This work is supported by 973 Program (2015CB857002), NSFC (11522324,11733004,11421303,11673015),
 and the Fundamental Research Funds for the
Central Universities. H.J.M. acknowledges the support of NSF AST-1517528.
The work is also supported by the Supercomputer Center
of the University of Science and Technology of China.

\bibliography{citations}

\end{document}